\documentclass[referee]{cjaa}           

\usepackage{graphicx}                   
\input{epsf.sty}                        
\input{psfig.sty}                       

\newcommand{\be}{\begin{equation}}
\newcommand{\ee}{\end{equation}}

\begin{document}

   \title{Short-living supermassive magnetar model for the early X-ray flares following short GRBs
}

   \volnopage{Vol.0 (200x) No.0, 000--000}      
   \setcounter{page}{1}          

   \author{Wei-Hong Gao\inst{1}\mailto{gaoweihong@njnu.edu.cn} \and Yi-Zhong Fan\inst{2,3,4}
}
   \offprints{W.H. Gao}                   

   \institute{Department of Physics and Institute of Theoretical Physics,
Nanjing Normal University, Nanjing, 210008, China\\
             \email{gaoweihong@njnu.edu.cn}
         \and The Racah Inst. of Physics, Hebrew University, Jerusalem 91904, Israel\\
         \and Purple Mountain Observatory, Chinese Academy of
Science, Nanjing 210008, China\\
       \and Lady Davis Fellow\\
          }
   \date{Received~~2006 month day; accepted~~2006~~month day}

\abstract{We suggest a short-living supermassive magnetar model to
account for the X-ray flares following short $\gamma-$ray bursts.
In this model, the central engine of the short $\gamma-$ray bursts
is a supermassive millisecond magnetar, that is formed in double
neutron stars coalescence. The X-ray flares are powered by the
dipole radiation of the magnetar. When the magnetar has lost a
significant part of its angular momentum, it collapses to a black
hole and the X-ray flares disappear abruptly.
}

   \authorrunning{W.H.Gao \& Y.Z. Fan}            
   \titlerunning{Short-living supermassive magnetar model for the early X-ray flares}

   \maketitle

%
%
\section{Introduction}           
\label{sect:Intro}
So far, the X-ray flares following short GRBs have been detected
in GRB 050709 and in GRB 050724 (Villasenor et al. 2005; Barthelmy
et al. 2005). In GRB 050709, a short GRB localized by HETE-II,
 X-ray flares occurred $\sim 10$ s and $\sim 16$ days after the
GRB (Fox et al. 2005). The details of the flares are unclear owing
to the rare data. Much better X-ray flare data is available for
GRB 050724, a short GRB localized by {\it Swift}. The  data has
been summarized in Barthelmy et al. (2005; see also Zhang et al.
2006). The X-ray telescope (XRT) observation starts at $\sim 79$ s
after the trigger. An extended flare-like epoch stops at $\sim
200$s after which the lightcurve decays rapidly (with a temporal
slope index $\sim -10$). A second, less energetic flare epoch
peaks at $\sim 300$ s, which is followed by another steep decay
with a slope $\sim -7$. In this work, we call the first and the
second flare epochs as the ``early X-ray flares". A third,
significant flare epoch starts at $\sim 2\times 10^4$ s, and the
decay slope after the peak is $\sim -2.8$. We call the third one
as the ``late X-ray flare".

The X-ray flares following short GRBs, like those detected in the
long GRB X-ray afterglows, may indicate the long activity of the
central engine (e.g., Barthelmy et al. 2005; cf.
 MacFadyen et al. 2006). But the actual mechanism for the long activity
 of the short GRB central engine is still unclear (Proga \& Zhang 2006;
 Perna et al. 2006; Dermer \& Atoyan 2006).  The difficult is that
 in the double neutron stars coalescence scenario (i.e., the leading model
 of the short GRBs; e.g., Eichler et al. 1989),
 the typical energy input episode
 is just in order of the duration of the short burst, provided that
 after the merger a black hole is formed (e.g., Rosswog \& Davis 2002; Lee et al.
 2004).

In this work, we suggest that the X-ray flares following short
GRBs could be understood if after the two neutron stars merger, a
millisecond rotating magnetized supermassive neutron star (SMNS)
instead a black hole, is formed.


\section{THE MODEL}
\label{sect:Model}
After the merger of two neutron stars, a supermassive neutron star
(SMNS) with a mass $\sim 2.5~M_\odot$, differentially rotating
with a period of $P\sim 1$ ms, can be formed (e.g., Klu\'znik \&
Ruderman 1998; Rossowog \& Davis 2002). The maximum stable mass of
a slowly rotating neutron star is $1.8-2.3~M_{\odot}$ (Akmal et
al. 1998), and the uniform rotation could increase these values by
at most $\sim 20\%$ (e.g. Cook et al. 1992, 1994, and references
therein). The SMNS therefore could survive before it has lost a
significant part of the angular momentum, if the state equation of
the nuclear material is stiff enough (cf. Shibata et al. 2006).

Though $B_o$, the initial surface magnetic field of the nascent
SMNS, may be just in order of $10^{12}$ G or lower\footnote{After
the submission of this work, several relevant papers appeared. Dai
et al. (2006) proposed a post-merger millisecond differential
rotating neutron star model to account for the X-ray flares
following short GRBs. Price \& Rosswog (2006) found out that in
their MHD simulation of the double neutron star coalescence, a
magnetar might be formed. Fan \& Xu (2006) suggested that the long
term X-ray flat segment detected in short GRB 051221A could be
well accounted for, provided that the central engine was a
magnetar. These last two findings are consistent with our model.},
much higher surface dipole magnetic field $B_{dip}$ may be
generated by several dynamo actions in a very short time. (i)
Currently, the Rossby number $R_o \leq 1$ (Rosswog, Ramirez-ruiz
\& Davis 2003), both the $\alpha^2$ and the $\alpha-\Omega$
dynamos could amplify the initial field effectively and
$B_{dip}\sim 10^{15}$ G is expected (Duncan \& Thompson 1992;
Thompson \& Duncan 1993). (ii) The convective dynamo can also
generate a very strong dipole filed (Duncan \& Thompson 1992).
(iii) If soon after the sudden formation of the SMNS, the
convective and hydrodynamical instabilities have been greatly
diminished. The
 magnetic field could be amplified by the linear
amplification process (Klu\'znik \& Ruderman 1998) and
$B_{dip}\sim 10^{15}$ G or stronger is still expected  in a
timescale $\sim 10~B_{o,12}^{-1}P_0$ seconds (Spruit 1999). Here
and throughout this text, the convention $Q_x=Q/10^x$ has been
adopted in cgs units.

When the surface magnetic field strength reaches $\geq 10^{15}$ G,
the differential rotation will be terminated in a very short
timescale by the magnetic braking (e.g., Shapiro 2000; Shibata et
al. 2006)

\begin{equation}
\tau\sim10^{2}(\frac{B_{dip}}{10^{15}G})^{-1}(\frac{R_s}{\rm 15
~km})^{-1/2}(\frac{M_s}{2.5M_{\odot}})^{1/2}~{\rm ms},
\label{eq:Gao1}
\end{equation}
where $R_s$ and $M_s$ are the radius and mass of the
differentially rotating SMNS, respectively. That means the
differentially rotating SMNS is estimated to evolve to a uniform
rotation profile on a timescale much shorter than the spindown
time of a uniformly rotating star(Eq. [3], derived below). So the
SMNS is mainly supported by the rapid uniform rotation rather than
the differential rotation.

The millisecond magnetar will lost their angular momentum quickly
through the dipole radiation and strong Poynting flux dominated
outflow is ejected. As a significant part of its angular momentum
has been lost, the SMNS is very likely to collapse to a black
hole. Before that time, the magnetic dissipation of the Poynting
flux dominated outflow may be able to power detectable X-ray
flares, which is of our interest. The dipole radiation luminosity
is (e.g., Usov 1994)
\begin{equation}
L\sim 3\times 10^{50}~{\rm
ergs~s^{-1}}~B_{dip,15}^2R_{s,6}^6\Omega_4^4, \label{eq:Gao2}
\end{equation}
where $\Omega$ is the angular velocity.

The corresponding spin-down timescale is
\begin{equation}
t_{sd}\sim 4\times10^{2}~{\rm
s}~\frac{j_{s}}{0.6}(\frac{M_{s}}{2.5M_{\odot}})^{2}
(\frac{R_{s}}{\rm 15~km})^{-6}\Omega_{4}^{-4} B_{dip,15}^{-2},
\label{eq:Gao3}
\end{equation}
where $j_s$ is the specific angular momentum. In this Letter we
focus on the case of the SMNS spinning down exclusively via a
magnetospheric wind, but it is also possible that significant
spindown can also occur through gravitational wave emission, such
as those driven by r-mode instabilities(Andersson 1998). A rough
estimation of this timescale is within a year (Vietri \& Stella
1998), much longer than the timescale through magnetic dipole
radiation. On the other hand, many aspects of the present
theoretical calculations regarding gravitational waves are
uncertain(Fryer and Woosley 2001). So in our Letter we just take
into account the electromagnetic spindown.

In the Poynting-flux dominated outflow, the X-ray flare emission
could be due to the dissipation of the magnetic field (Usov 1994;
Thompson 1994) or internal shocks with magnetization (Fan et al.
2004). For illustration, here we take the global MHD condition
breakdown model to calculate the emission. By comparing with the
pair density ($\propto r^{-2}$, $r$ is the radial distance from
the central source) and the density required for co-rotation
($\propto r^{-1}$ beyond the light cylinder of the compact
object), one can estimate the radius at which the MHD condition
breaks down, which reads (Usov 1994; Zhang \& M\'esz\'aros 2002)
\begin{equation}
r_{MHD}\sim2\times10^{16}L_{50}^{1/2} \sigma_1^{-1} {t_{v,m}}_{-3}
\Gamma_{2}^{-1}~{\rm cm},
\end{equation} where $\sigma$ is the ratio of the magnetic energy
flux to the particle energy flux, $\Gamma$ is the bulk Lorentz
factor of the outflow, $t_{v,m}$ is the minimum variability
timescale of the central engine. Beyond this radius, intense
electromagnetic waves are generated and outflowing particles are
accelerated  (e.g. Usov 1994). Such a significant magnetic
dissipation process converts the electromagnetic energy into
radiation. The radiation should be delayed in a timescale
(relative to the initial hard $\gamma-$ray spike)
\begin{equation}
\tau_{delay}\sim {r_{MHD} \over 2 \Gamma^2 c}=33 ~{\rm
s}~L_{50}^{1/2} \sigma_1^{-1} {t_{v,m}}_{-3} \Gamma_{2}^{-3}~{\rm
cm},
\end{equation}
which matches the observation of GRB 050709 and GRB 050724.

At $r_{MHD}$, the corresponding synchrotron radiation frequency
can be estimated as (Fan et al. 2005)

\begin{equation}
\nu_{m,MHD}\sim6\times10^{16}\sigma_1^{3}C_{p}^{2} \Gamma_{2}
{t_{v,m}}_{-3}(1+z)^{-1}~{\rm Hz},
\end{equation}
where
$C_{p}\equiv(\frac{\epsilon_{e}}{0.5})[\frac{13(p-2)}{3(p-1)}]$,
$\epsilon_{e}$ is the fraction of the dissipated comoving magnetic
field energy converted to to the comoving kinetic energy of the
electrons, and the accelerated electrons distribute as a single
power-law $dn/d\gamma_{e}\propto\gamma_{e}^{-p}$. So most energy
is radiated in the soft X-ray band.

For the magnetic-dominated $e^{+}e^{-^{}}$ plasma, the diamagnetic
relativistic pulse accelerator (DRPA) mechanism can convert most
of the initial magnetic energy into the ultrarelativistic energy
of a fraction of the surface particles. In the numerical
simulation of such a plasma, Liang \& Nishimura (2004) discovered
that the plasma pulse bifurcated repeatedly, leading to a complex,
multipeak structure at late times. {\it So the flares from the
magnetic dissipation can show repeated, multiple structures.}
Alternatively, the multiple flares may suggest that the magnetic
dissipation takes place just locally rather than globally. The
emission from different dissipation region arrives at different
time and thus gives rise to multiple flares (Giannios 2006).

Our model is based on the double neutron stars merger model for
short gamma-ray bursts (e.g. Eichler et al. 1989), which seems to
be supported by the current host galaxy and afterglow observations
(Barthelmy et al. 2005; Fox et al. 2005; Gehrels et al. 2005). The
double neutron stars merger model is also consistent with the rate
and the luminosity function of short GRBs detected by HETE-2/{\it
Swift} (Piran \& Guetta 2006, and the references therein). The
neutron star-black hole merger model is an important alterative.
But in the black hole and neutron star merger scenario, it is very
hard to produce the X-ray flares. {\it Therefore, if X-ray flares
following a short GRB have been detected and the short GRB is
found to be outside of the galaxy, the double neutron star merger
model is strongly supported (Note that just double neutron star
and black hole-neutron star merger can occur outside of the host
galaxy, see Fryer et al. [1999] for details).} The X-ray flares
then could play an additional role. When the double neutron star
merger occurring outside of their host galaxy, the X-ray afterglow
emission seems to be very weak and may be undetectable for the XRT
in a long term, as in the case of GRB 050709 (Gehrels et al.
2005). However,  the energetic X-ray flares provide us the chance
to  get a much better location.

\section{DISCUSSION}

{\em Swift} XRT has revealed a new, rich, and unexpected
phenomenology of early X-ray afterglows observations. The most
important one may be the energetic flares observed hundreds to
thousands of seconds after the initial burst signal in both long
and short GRBs (e.g. Burrows et al. 2005; Barthelmy et al. 2005).
All these X-ray flares can be well interpreted by the ``inner
energy dissipation model", for example, the late internal shock
model (Fan \& Wei 2005; Zhang et al. 2006) and/or the late
magnetic energy dissipation model (Fan et al. 2005). In this
model, the long activity of the central engine is needed.
However, a proper model for the long activity of the short GRB is
still unavailable.

In this Letter, we suggest a magnetized central engine model for
the X-ray flares in short GRBs. We take the popular two neutron
stars coalescence model (e.g., Eichler et al. 1989) but to assume
that after the merger, a differential rotating neutron star rather
than a black hole, is formed. Soon after its sudden formation, the
initial magnetic field has been amplified significantly by several
possible dynamos (e.g., Duncan \& Thompson 1992; Thompson \&
Duncan 1993; Klu\'zniak \& Ruderman 1998; Spruit 1999), and the
dipole magnetic field could be as strong as $\sim 10^{15}$ G. The
differential rotation of the SMNS has been terminated by the
magnetic braking. The early X-ray flares are powered by the dipole
radiation of the nascent neutron star. They turn off when the
supermassive neutron star collapses to a black hole. If this
scenario is correct, X-ray flares much more energetic than that
detected in GRB 050724 may be detectable in the coming months and
years by {\it Swift}. The other speculation is that some short
GRBs with X-ray flares would occur outside of their host galaxies.

The outflow powering the X-ray flares are Poynting-flux dominated,
so the emission should be linearly polarized, as suggested by Fan
et al. (2005). In this model, no strong MeV-GeV photons
accompanying the X-ray flare (due to the synchrotron self inverse
Compoton effect) are expected, in contrast to the baryon-rich
internal shock model (see Wei et al. 2006 for a primary
suggestion).

\section*{Acknowledgments}
Y. Z. Fan is supported by US-Israel BSF, the National Natural
Science Foundation (grants 10225314 and 10233010) of China, and
the National 973 Project on Fundamental Researches of China
(NKBRSF G19990754).

\end{document}